\documentclass[journal,twoside,print]{ieeecolor}
\usepackage{booktabs}
\usepackage{float}
\usepackage{tmi}
\usepackage{cite}
\usepackage{amsmath,amssymb,amsfonts}
\usepackage{algorithmic}
\usepackage[]{graphicx}
\usepackage{textcomp}
\usepackage{stmaryrd}  
\DeclareUnicodeCharacter{2212}{-}

\def\BibTeX{{\rm B\kern-.05em{\sc i\kern-.025em b}\kern-.08em
    T\kern-.1667em\lower.7ex\hbox{E}\kern-.125emX}}
    
\begin{document}
\title{Neurovascular Segmentation in sOCT with Deep Learning and Synthetic Training Data.}


\author{Etienne Chollet$^\dagger$, Ya\"el Balbastre$^\dagger$, Chiara Mauri, Caroline Magnain, Bruce Fischl$^\ddagger$, and Hui Wang$^\ddagger$
\thanks{$^\dagger$The first two authors contributed equally to this work. $^\ddagger$The last two authors contributed equally to this work.  
This work was supported in part by the National Institutes of Health (1R01 AG070988-01,  1RF1MH123195-01, K99 HD101553, U01 AG052564, R56 AG064027, R01 AG064027, R01 AG016495, U01 MH117023, P41 EB015896, R01 EB023281, R01 EB019956, R01 NS0525851, R21 NS072652, R01 NS083534, U01 NS086625, U24 NS10059103, R01 NS105820, S10 RR023401, S10 RR019307,  S10 RR023043), the Royal Society (NIF{\textbackslash}R1{\textbackslash}232460) and the Chan-Zuckerberg Initiative DAF (2019-198101, 2021-244261).}
\thanks{Etienne Chollet is with the Athinoula A. Martinos Center for Biomedical Imaging, Massachusetts General Hospital, Charlestown, MA 02129 USA (e-mail: echollet@mgh.harvard.edu).}
\thanks{Ya\"el Balbastre was with the Athinoula A. Martinos Center for Biomedical Imaging, Massachusetts General Hospital, Charlestown, MA 02129 USA, and also with the Department of Radiology, Harvard Medical School, Boston, MA 02115 USA. He is now with the Department of Experimental Psychology, University College London, London, UK (e-mail: y.balbastre@ucl.ac.uk).}
\thanks{Caroline Magnain, Chiara Mauri and Hui Wang are with the Athinoula A. Martinos Center for Biomedical Imaging, Massachusetts General Hospital, Charlestown, MA 02129 USA, and also with the Department of Radiology, Harvard Medical School, Boston, MA 02115 USA (e-mail: cmagnain@mgh.harvard.edu; cmauri@mgh.harvard.edu;  hwang47@mgh.harvard.edu).}
\thanks{Bruce Fischl is with Athinoula A. Martinos Center for Biomedical Imaging, Massachusetts General Hospital, Charlestown, MA 02129 USA, also with the Department of Radiology, Harvard Medical School, Boston, MA 02115 USA, and also with the Computer Science and Artificial Intelligence Laboratory, MIT, Cambridge, MA 02139 USA (e-mail: bfischl@mgh.harvard.edu).}}

\maketitle
\thispagestyle{empty}

\begin{abstract}
Microvascular anatomy is known to be involved in various neurological disorders. However, understanding these disorders is hindered by the lack of imaging modalities capable of capturing the comprehensive three-dimensional vascular network structure at microscopic resolution. With a lateral resolution of $<=$20 {\textmu}m and ability to reconstruct large tissue blocks up to tens of cubic centimeters, serial-section optical coherence tomography (sOCT) is well suited for this task.
This method uses intrinsic optical properties to visualize the vessels and therefore does not possess a specific contrast, which complicates the extraction of accurate vascular models. The performance of traditional vessel segmentation methods is heavily degraded in the presence of substantial noise and imaging artifacts and is sensitive to domain shifts, while convolutional neural networks (CNNs) require extensive labeled data and are also sensitive the precise intensity characteristics of the data that they are trained on. Building on the emerging field of synthesis-based training, this study demonstrates a synthesis engine for neurovascular segmentation in sOCT images. Characterized by minimal priors and high variance sampling, our highly generalizable method tested on five distinct sOCT acquisitions eliminates the need for manual annotations while attaining human-level precision. Our approach comprises two phases: label synthesis and label-to-image transformation. We demonstrate the efficacy of the former by comparing it to several more realistic sets of training labels, and the latter by an ablation study of synthetic noise and artifact models. 
\end{abstract}


\section{Introduction}

\begin{figure*}[!t]
    \centering
    \includegraphics[width=0.9\textwidth]{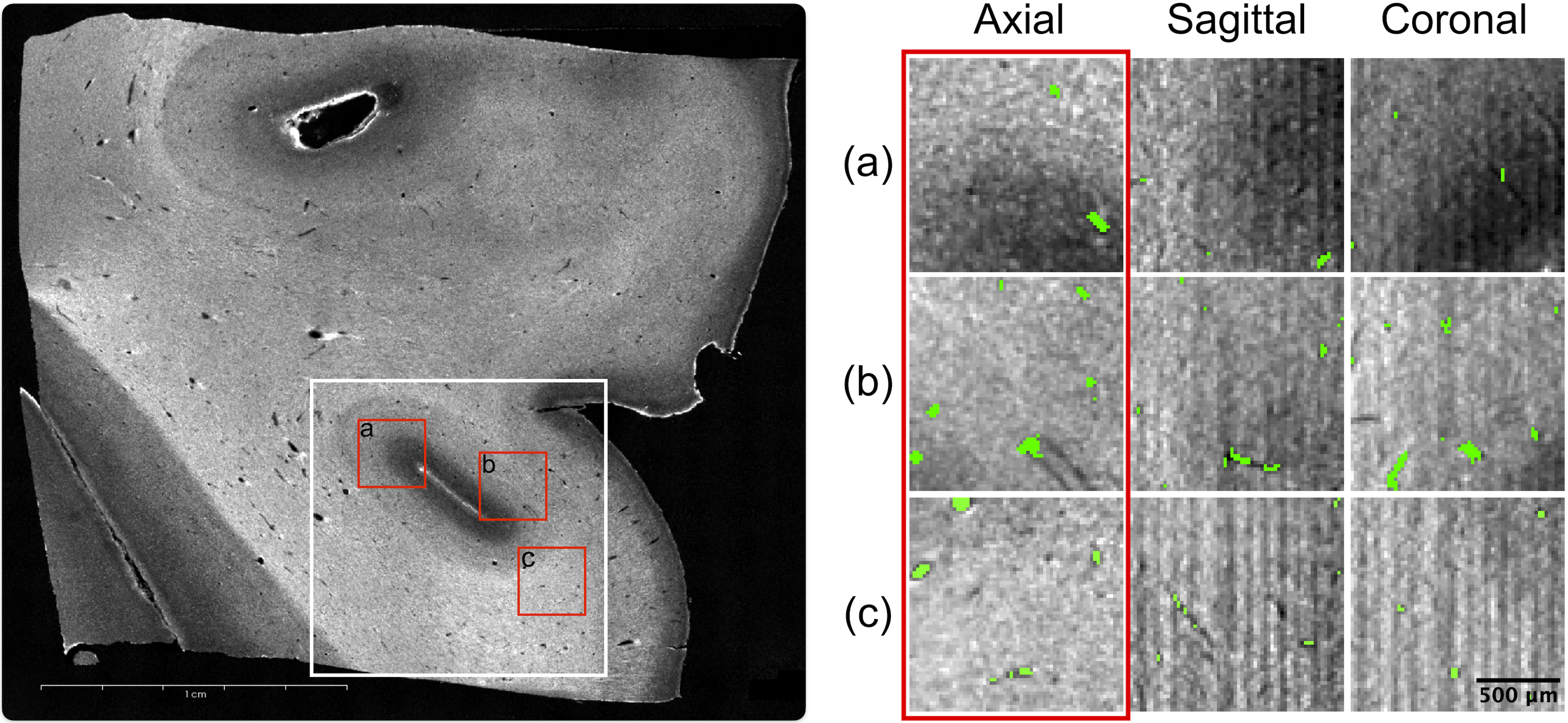}
    \caption{Visualization of sOCT data and manual segmentation of the vasculature. Left panel: A slice overview of an sOCT acquisition with regions of interest (a, b, c) outlined in red, each showing different anatomical features. Scale bar (1cm) shown in white in bottom left corner. Right panels: detailed manual segmentation for the corresponding regions viewed in axial, sagittal, and coronal planes, overlaid with segmented vessels (green). Scale bar (500 {\textmu}m) shown in black in bottom right corner.}
    \label{fig:I46_ref}
\end{figure*}

\subsection{Vasculature Segmentation in Biomedical Imaging}
Vasculature plays a pivotal role in the characterization of diverse medical conditions. The traditional “gold standard” for segmenting vessels in biomedical imaging involves manual annotation by highly skilled professionals. This method, however, demands extensive human expertise and labor, making it prohibitively expensive for routine use. Alternative, less expensive approaches rely on the hand-crafted priors of Hessian-based filters (Frangi, Sato, Meijering)\cite{frangi1998multiscale,yang2022volumetric}, morphological operations\cite{zana2001segmentation}, and region growing techniques\cite{dokladal1999liver}. Nonetheless crucial in contemporary biomedical imaging, these methods often degrade when vessels display atypical morphologies, cross tissue boundaries, or have low contrast to noise ratios (CNRs) \cite{yang2022volumetric}. In many cases, it has been demonstrated that multi-scale hessian-based filters produce artifactual structures that closely resemble the vasculature as well as blurring and distorting critical features such as vessel radius \cite{yousefi2015segmentation,longo2020assessment}. The extent of distortion is directly influenced by the range of scales selected for the "vesselness" filters and ultimately leads to significant inaccuracies in the extraction of vascular anatomy. Moreover, these algorithms exhibit high sensitivity to noise, and do not generalize well across different imaging modalities.
Convolutional neural networks (CNNs) have demonstrated robustness against the inconsistencies prevalent in traditional vascular segmentation algorithms \cite{goni2022brain}. This proficiency stems from their capacity to learn spatial hierarchies of features through convolutions, thereby enabling detailed segmentation tasks such as delineating both small (microscale) and large (mesoscale) vascular structures \cite{moccia2018blood,meijs2018artery}. As a result, this method significantly outperforms traditional automated methods \cite{jiang2023noise}. Recent advances in learning-based vascular segmentation include losses robust to imbalance between the foreground and background classes \cite{tetteh2020deepvesselnet}, losses that promote topological correctness \cite{araujo2019deep,kong2020learning,shit2021cldice,cheng2021joint,li20213d}, the use of limited or two-dimensional annotations \cite{kozinski2018learning}, uncertainty modeling \cite{rempfler2017uncertainty, hu2023learning, gupta2024topology}, and models trained across many different imaging modalities \cite{holroyd2023tube}. However, a significant challenge persists: the efficacy of CNNs relies heavily on the quantity and quality of training data. This dependency underscores a critical gap in our ability to effectively analyze vascular features, particularly in the context of data scarcity and novel tasks.

\subsection{Serial sectioning Optical Coherence Tomography}

Serial-sectioning Optical Coherence Tomography (sOCT) leverages tissue's intrinsic optical properties to efficiently acquire volumetric data without labeling \cite{wang2018psoct} far more rapidly than is currently possible with traditional microscopy techniques. By depth-resolving the back-scattered intensity of impinged light, sOCT is capable of extracting many different anatomical features including the neural parenchyma, high-scattering myelinated fibers, and low-scattering blood vessels. At the finest scale, sOCT has been shown to identify small arterioles and venules down to a diameter of 20 {\textmu}m \cite{yang2022volumetric}. This modality not only captures volumetric data efficiently—one acquisition encodes the information across several hundred micrometers in depth—but also integrates a tissue slicer into the imaging workflow, enabling the reconstruction of large tissue blocks up to tens of cubic centimeters. Crucially, sOCT minimizes tissue distortion by imaging directly on the block face \emph{before} sectioning is performed. With a lateral resolution of $\sim$1-20 {\textmu}m, sOCT is exceptionally well-suited for resolving fine vascular structures, as depicted in Fig. \ref{fig:I46_ref}. This capability marks sOCT as a useful tool for scientists seeking a high-throughput, high-resolution imaging of 3D vascular networks.

While the use of sOCT for vascular imaging holds considerable promise, several challenges complicate the effective extraction of vessels from sOCT data. Despite its advantageous label-free imaging capabilities, sOCT lacks the specificity associated with fluorescent-based methods that target a diverse set of biological markers. Thus, the CNR of vessels relative to surrounding tissue is low compared to contrast-enhanced images that arise from targeted stains. This issue is exacerbated by the pervasive presence of granular gamma-distributed speckle noise—attributable to the use of a low-temporal coherent light source—which often matches the size of the small structures of interest\cite{kirillin2014speckle}. sOCT images also exhibit intensity decay along the depth of acquired sections due to the attenuation of light as it propagates through tissues. This attenuation varies significantly with tissue type, making direct correction particularly challenging. The artifact is only exacerbated when stitching multiple sections together to reconstruct volumetric data, resulting in many slab-wise nonuniformities in intensity (banding).

\begin{figure*}[!t]
    \centering
    \includegraphics[width=\textwidth]{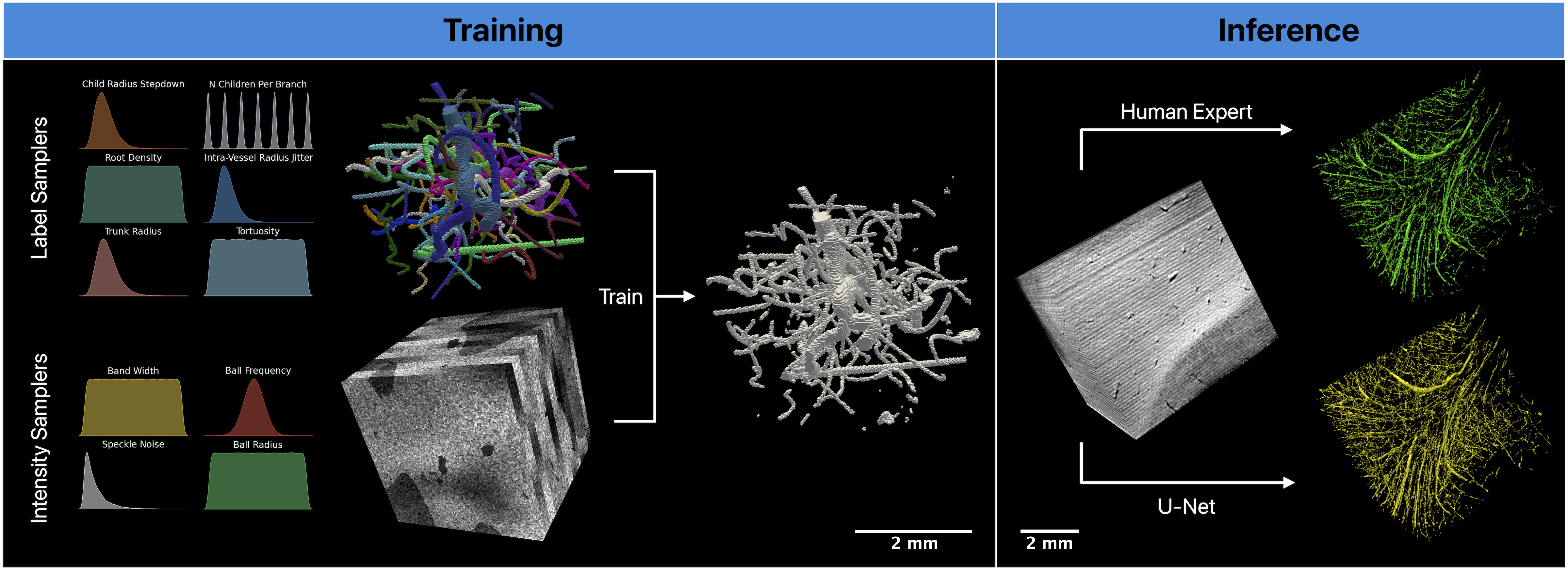}
    \caption{Overview of proposed method. Labels and intensity textures are synthesized to resemble volumetric sOCT data which is used to train a U-Net in binary vessel segmentation. We then use these models for large-scale prediction and compare with expert labelers.}
    \label{fig:graphical_abstract}
\end{figure*}

\subsection{Synthetic Data in Machine Learning}
In the evolving landscape of machine learning in medical imaging, synthetic data generation stands out as an attractive approach that is designed to overcome limitations associated with scarce or inconsistent training datasets. The family of \texttt{Synth} techniques (\texttt{SynthSeg}\cite{billot2023synthseg}, \texttt{SynthMorph}\cite{hoffmann2021synthmorph}, \texttt{SynthSR}\cite{iglesias2021joint,iglesias2023synthsr}, \texttt{SynthStrip}\cite{hoopes2022synthstrip}) exemplify this trend. These techniques facilitate the creation of effectively infinite, complex, and perfectly annotated imaging data without the need for extensive manual annotations of the images acquired in the target domain.

While traditional data augmentation techniques such as geometric transformations and intensity or contrast manipulations increase the variance within training datasets, they often fail to adequately represent outlying features that are crucial for robust model training. Moreover, approaches such as generative adversarial networks (GANs) tend to only replicate existing patterns, potentially overlooking the unique, less common features necessary for a rich learning environment \cite{skandarani2023gans}.

Another appealing option in the synthetic route is to train models on datasets whose vascular trees have been obtained via physiology-based simulations \cite{schneider2012tissue} or constrained constructive optimization (CCO) \cite{tetteh2020deepvesselnet}. These techniques can be used in an application-specific manner to adapt the generated data to a particular vascular segmentation task such as segmenting vessels from optical coherence tomography-angiography (OCTA) data \cite{menten2022physiology}. These techniques nonetheless encode strong priors that may degrade the accuracy of results when the priors do not align with observed vascular morphology.

To circumvent these shortcomings, synthetic datasets such as those produced via the \texttt{Synth} framework are engineered to synthesize a wide array of data, specifically including a wider distribution than is expected to be seen at inference time, fostering the creation of models that are both versatile and robust. These techniques generate synthetic brain images in perfect register with their semantic labels, since the images are generated from the labels, enabling the development of algorithms that are resilient across diverse conditions. SynthMorph further demonstrated that informative representations can be obtained with purely synthetic geometrical labels. This idea was later leveraged and extended in AnyStar \cite{dey2024anystar}, a cell segmentation model trained without any real labels or images.

\subsection{Contribution}

Our solution to the problem of vascular segmentation in sOCT data relies on a synthesis strategy in line with previous work \cite{billot2023synthseg,hoffmann2021synthmorph,dey2024anystar}. In keeping with the theme of high variance sampling and \textit{unrealistic} domain randomization exemplified by these studies, we create a rich learning environment and eliminate the need for real imaging data and expensive manual labels. Importantly, we deviate from research that leverage realistic synthetic vascular networks \cite{tetteh2020deepvesselnet, menten2022physiology}, in that we follow a purely geometrical argument. We make the assumption, akin to vesselness filters, that a structure is a vessel if and only if it is long and thin. During training, a variety of shapes are displayed, some of them tubular. Because our training data only contains very generic characteristics about ``vessels'', the resulting networks are much more robust to domain shifts than those trained with alternative strategies.

Specifically, we demonstrate the following contributions:

\begin{itemize}{
    \item \textbf{Spline Based Synthetic Label Engine:} We propose a synthesis engine that uses unrealistic spline-based geometries to generate a wide array of vessel-like structures, far exceeding the complexity observed in real-world data.

    \item \textbf{Synthetic Image Engine:} We model a superset of domain-randomized intensity and textural artifacts found in sOCT.
    
    \item \textbf{Accurate \& Precise Vasculature Segmentation in sOCT Data:} Our method achieves segmentation accuracy and precision statistically comparable to human experts across four distinct imaging conditions.
    
}\end{itemize}

\section{Proposed Method}

We employ domain-randomized synthesis to generate structured labels, textures, and intensity volumes (summarized in Fig. \ref{fig:graphical_abstract}). Collectively, these are parameterized by sampled quantities such as number of unique classes, vessel geometry, and image artifacts. All parameters are sampled randomly from one of four statistical distributions:
\begin{itemize}
    \item $\mathcal{N}(\mu, \sigma^2)$: the normal distribution, with mean $\mu$ and variance $\sigma^2$;
    \item $\mathcal{LN}(\mu, \sigma^2)$:  the log-normal distribution, where $\mu$ and $\sigma^2$ are the mean and variance of the logarithm of the data;
    \item $\mathcal{U}(a, b)$: the continuous uniform distribution on $[a, b]$;
    \item $\mathcal{U}_{\mathbb{Z}}(a, b)$: the discrete uniform distribution on $\llbracket a, b \rrbracket$.
\end{itemize}
We sample with high variance to include many "fringe" or "outlier" cases, which are vital for robust training of CNNs. As such, we aim to synthesize data that is a superset of what we expect from sOCT acquisitions. That is, containing the real distribution but with much more variance.

\subsection{Synthetic Labels}
Our method for synthesizing vascular labels utilizes spline-based computational geometry to not only approximate the vessels seen by sOCT images but purposefully expand the geometric complexities of vascular systems within a three-dimensional space. We generate arborescence-rooted trees modeled as bifurcating cubic splines to resemble the branched, curvilinear structures of vascular manifolds. We initialize an empty tensor ($128^3$ voxels) with an isotropic resolution of 20 {\textmu}m/voxel. The number of vascular trees in a given volume sampled from $\mathcal{U}(0.1, 0.2)$ trees/mm$^3$ (approximately 2-3 trees per volume), the number of offspring per spline from $\mathcal{U}_{\mathbb{Z}}(1, 7)$, and the maximum tree depth from $\mathcal{U}_{\mathbb{Z}}(1, 7)$. At each level, the endpoints of each spline are sampled first, and intermediate control points are jittered to match a target random tortuosity, sampled in $\mathcal{U}(1, 5)$. The radius of root splines is sampled from $\mathcal{LN}(-1, 0.017)$ {\textmu}m, and the radius of any branching spline is obtained by multiplying the preceding radius by a random factor sampled from $\mathcal{LN}\left(-\frac{1}{3}, 0.064\right)$. Furthermore, the radius of each spline is allowed to vary along its length and is also encoded by cubic splines, with a fluctuation factor sampled from $\mathcal{LN}(-0.0085, 0.017)$. Once all trees are sampled,  they are rasterized on the $128^3$ lattice. This involves computing the distance from any voxel to its nearest point on a spline, and checking whether it is smaller than the spline radius at that point. Distances are computed using a succession of discrete, quadratic and Gauss-Newton optimization loops \cite{wang2002robust}. Rasterization is the most computationally intensive step of our synthesis process, therefore a large number of patches are generated offline. However, since each individual spline is given a unique label, additional on-the-fly randomization is possible--such as random removal of some of the branches.

\begin{figure}[ht!]
    \centering
    \centering
    \includegraphics[width=\columnwidth]{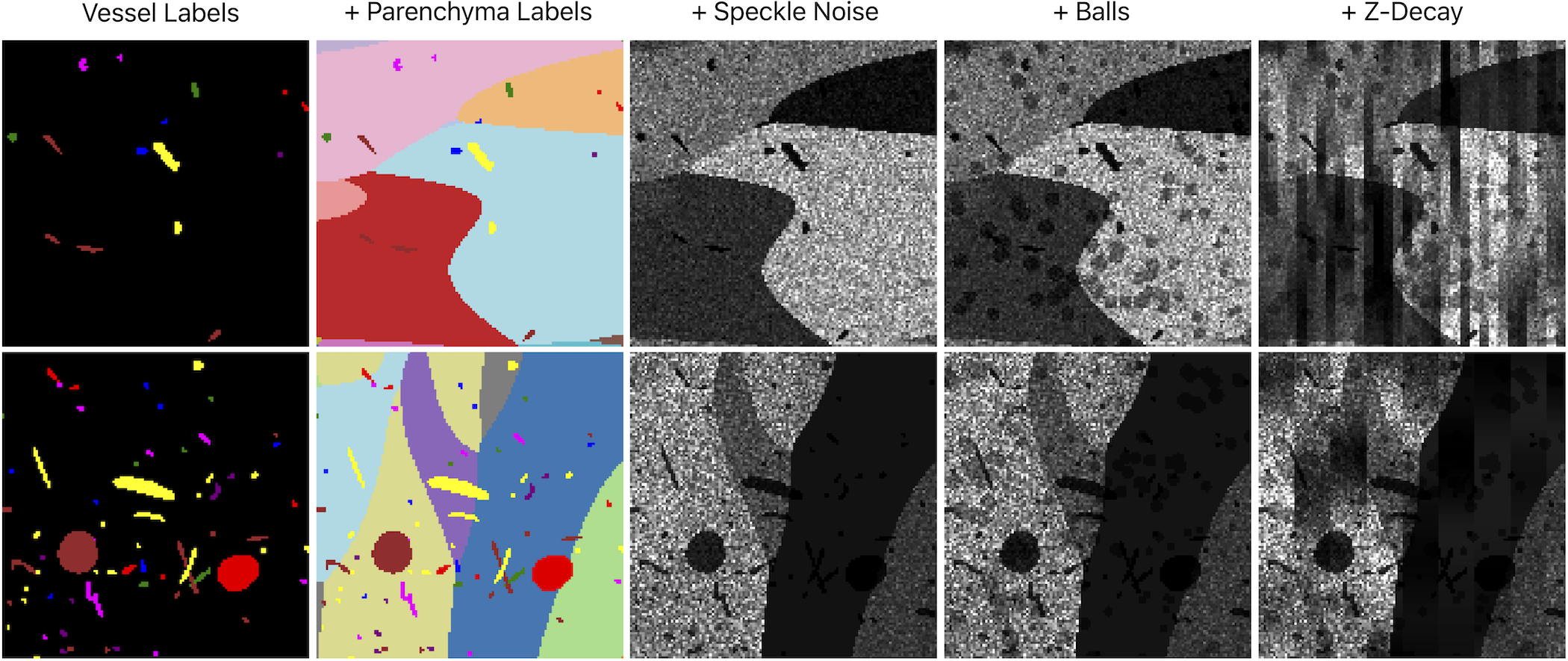}
    \caption{Non-exhaustive demonstration of texture mapping procedure for three separate instances of combined label maps.}
    \label{fig:synth_workflow}
    \includegraphics[width=\columnwidth]{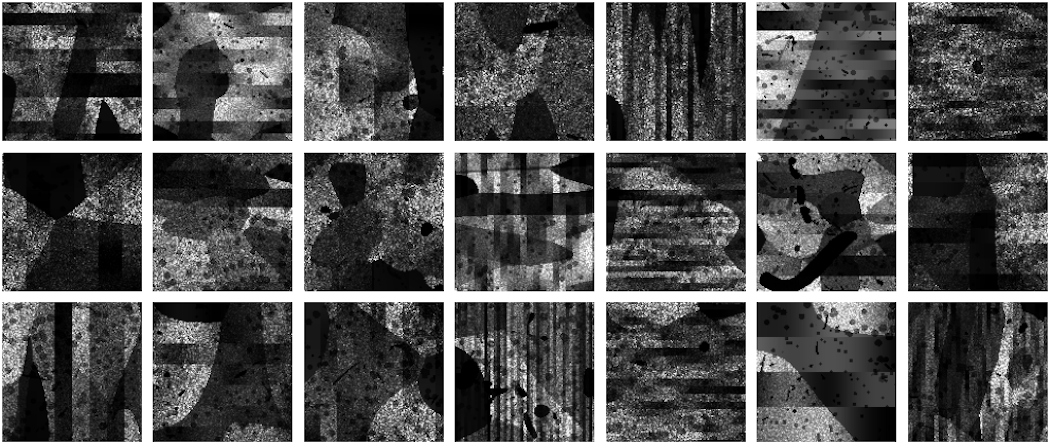}
     \caption{Random slices of synthetic data generated by our engine. Twenty-one examples show the wide range of textures, contrast gradients, and structures that create a rich training dataset.}
    \label{fig:synth_samples}
\end{figure}

\subsection{Synthetic Images}

Labels are converted into intensity images on-the-fly, such that no two images are unlikely to ever be the same and networks are trained on a virtually infinite dataset. Spatially continuous neural parenchyma labels are synthesized by sampling a set of smooth probability maps (encoded by three-dimensional cubic splines) and keeping the index of the map with maximum probability. The number of probability maps (and therefore the number of unique labels) is sampled from $\mathcal{U}_{\mathbb{Z}}(2, 10)$, and the number of spline control points along each of the three dimensions is sampled from $\mathcal{U}_{\mathbb{Z}}(3, 10)$. Each parenchyma label is assigned an intensity sampled from $ \mathcal{N}(i, 0.04)$ where $i$ represents the unique id of each integer label. The result is normalized to [0, 1]. Intra-vessel intensities and textures are added using a separate set of smooth label maps, to which Gaussian mixture transforms (with each mean sampled from $\mathcal{U}(0.7, 1)$, and a fixed variance $\sigma^2=0.64$) are applied. The result is normalized to [0, 0.5]. We fuse this with parenchyma intensities (normalized to [0, 1]) via element-wise multiplication to ensure the vessels are darker than the surrounding parenchyma, as shown in sOCT images.
To replicate the contamination of speckle noise in sOCT and modulate the CNR, we sample multiplicative noise from a Gamma distribution with a standard-deviation sampled from $\mathcal{U}(0.2, 0.8)$ and the mean fixed at one. The banding artifacts caused by stitching multiple sections are simulated by introducing multiple bands of slab-wise nonuniformities in intensity (width: $ \mathcal{U}_{\mathbb{Z}}(2, 32)$ voxels) in a single plane throughout the volume. Additionally, random spherical objects (radius: $\mathcal{U}_{\mathbb{Z}}(2, 8)$ voxels, frequency: $\mathcal{U}(10^{-3}, 10^{-5})$, intensity: $\mathcal{U}(0.1, 2)$) are added to address challenges met during the development of this method, as these objects in real sOCT data often led to many false positives in vessel segmentation. The intensity synthesis process
is graphically presented in Fig. \ref{fig:synth_workflow}. In all, the synthesized intensity creates a diverse set of training examples, as shown in Fig. \ref{fig:synth_samples}.

\subsection{Deep Learning}
We used a U-Net \cite{ronneberger2015u} with two residual blocks per layer \cite{he2016identity}, each made of a convolution, ReLU and instance normalization. The input layer was sized at $128^3$, with the base number of features (32) doubling for each of the four levels resulting in $7.4 \cdot 10^6$ trainable parameters. The network was trained on entirely synthetic data with a Dice loss on the foreground labels:
\begin{equation}
    \mathcal{L}_{\text{Dice}} = 1 - \frac{2~\mathbf{y}^\mathrm{T} \hat{\mathbf{y}} + \varepsilon}{\mathbf{y}^\mathrm{T}\mathbf{y} + \hat{\mathbf{y}}^\mathrm{T}\hat{\mathbf{y}} + \varepsilon} ~,
\end{equation}
where $\mathbf{y}$ indicates the ground truth label map, $\hat{\mathbf{y}}$ the predicted probability map, and $\varepsilon$ is a stabilizing constant, which we set to the number of voxels in the patch ($\varepsilon=\mathbf{1}^\mathrm{T}\mathbf{1}$). The model was trained for a total of 100,000 steps using the Adam optimizer ($\beta_1=0.9, \beta_2=0.999, \epsilon=10^{-8}$). The learning rate was linearly warmed up from $10^{-10}$ to $10^{-2}$ over a period of 2,000 steps, held constant at $10^{-2}$ for the majority of training, then linearly cooled down from $10^{-2}$ to $10^{-6}$ starting at the 80,000$^{th}$ step. Each model was trained three times with different, random weight initializations.

\subsection{Prediction}
To predict on a large sOCT volume, a sliding window of 128$^3$ voxels with 32-voxel steps (in all dimensions) is used to average individual patch predictions. We weight individual patch predictions using a sine function on $\left[\frac{\pi}{8}, \frac{7\pi}{8}\right]$, centered at index 64 in each dimension, to attenuate predictions near the edges (where input context is scarce). This results in a 64x averaging of each voxel in the input image.

\subsection{Evaluation Metrics}
To assess the accuracy of our segmentation method, we compare model predictions to expert annotations using the Dice-Sørensen coefficient (DSC), the false positives rate (FPR) and false negatives rate (FNR), defined as follow:
\begin{equation*}
    \textstyle
    \text{DSC} = 
    \frac{2\text{TP}}{2\text{TP}+\text{FP}+\text{FN}}
    ,~~~
    \text{FPR} = 
    \frac{\text{FP}}{\text{TN}+\text{FP}}
    ,~~~
    \text{FNR} = 
    \frac{\text{FN}}{\text{TP}+\text{FN}}
    ,
\end{equation*}
where TP, TN, FP, FN denote the number of true positives, true negatives, false positives, and false negatives, respectively.

\section{Experiments}

\subsection*{Brain samples and sOCT acquisition}
Three human brains were obtained from the Massachusetts General Hospital (MGH) Autopsy Suite. The subjects were neurologically normal prior to death. The brains were fixed by immersion in 10 $\%$ formalin for at least two months. The ex-vivo imaging procedures are approved by the Institutional Review Board of the MGH. A sample of primary somatosensory cortex ($3.4 \times 2.9 \times 1.1$ cm$^3$) was obtained from one brain and imaged by a 1300 nm spectral domain sOCT system, with an in-plane resolution of 10 {\textmu}m and an axial resolution of 3.5 {\textmu}m \cite{magnain2014blockface}. Both the frontal and occipital cortices were obtained from the other two brains and imaged by a different 1300 nm polarization sensitive sOCT system \cite{liu2022refractive}, with a lateral resolution of 10 {\textmu}m and axial resolution of 5 {\textmu}m. Image reconstruction has been elaborated in previous studies \cite{wang2018psoct}. All the volumetric reconstructions yield an isotropic resolution of 20 {\textmu}m. Vessels were manually labeled in a subset volume ($6 \times 6 \times 6$ mm$^3$) by an expert, as well as in small patches ($1.3 \times 1.3 \times 1.3$ mm$^3$) of the frontal and occipital cortices by two labelers. 

\subsection*{Experiment A. Ablation study}

Our proposed method includes 100,000 high-variance training examples incorporating the noise models described in II.B. The following experiment illustrates the detrimental effect of removing different noise and artifact augmentations from the synthesized sOCT images.

\subsubsection{Training Data}
Eight distinct training sets (A-I) were synthesized by permutational ablation of two image artifact models: spheres and intensity-banding, and one vessel-specific model: intra-vessel texture, from our proposed texture mapping method.

\subsubsection{Validation Data}
The development dataset on which we optimized the original model architecture and hyper-parameters consists of a $301^3$ voxel patch (white box in Fig. \ref{fig:I46_ref}) of primary somatosensory cortex, annotated for vasculature by an expert (green labels in right panel). A tissue mask was also made by the same expert to exclude non-tissue (background) regions.

\subsubsection{Setup}
We conducted a comprehensive training regimen with 100,000 training examples involving a single U-Net architecture, replicated three times across the eight distinct training-data-synthesis conditions (A-I). Condition A represents the full model whose hyper-parameters were optimized on the validation data, while conditions B-H represent different  combinations of ablated noise models. Model inferences are compared with expert labels using DSC, FPR, and FNR.

\begin{figure*}
    \centering
    \includegraphics[width=0.97\textwidth]{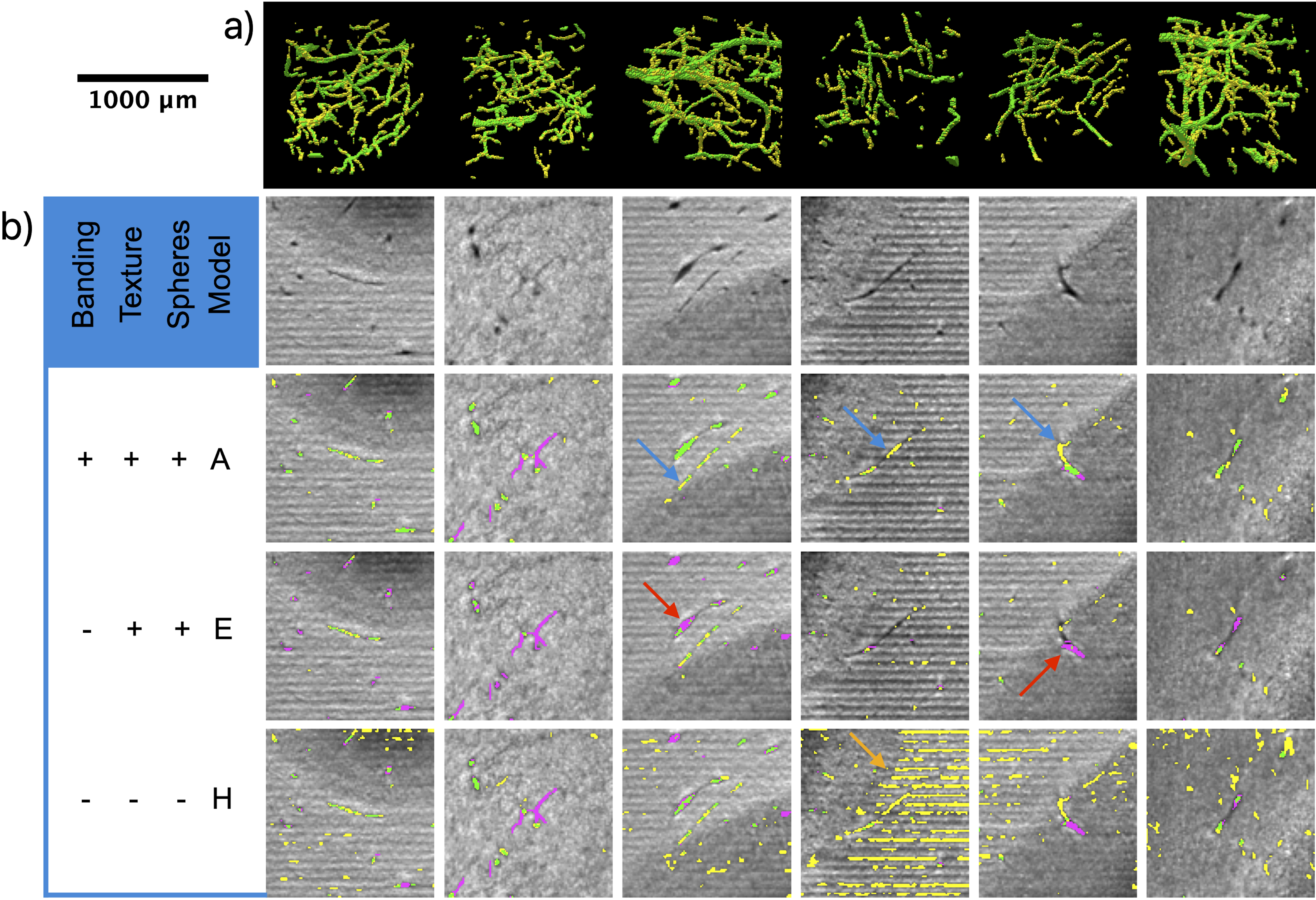}
    \caption{Qualitative comparison of model performance in different regions of normal control samples taken from the human somatosensory cortex. Subfigure (a) illustrates 3D renderings of TPs (green) and FPs (yellow) from the prediction of our proposed model (A). Subfigure (b) shows the same information but as 2d sections and with the addition of FNs (magenta). Scale bar (1000 {\textmu}m) shown in black in upper left corner.}
    \label{fig:ablation_images}
\end{figure*}

\subsubsection{Results}

\begin{figure}[t!]    
    \centering
    \includegraphics[width=\columnwidth]{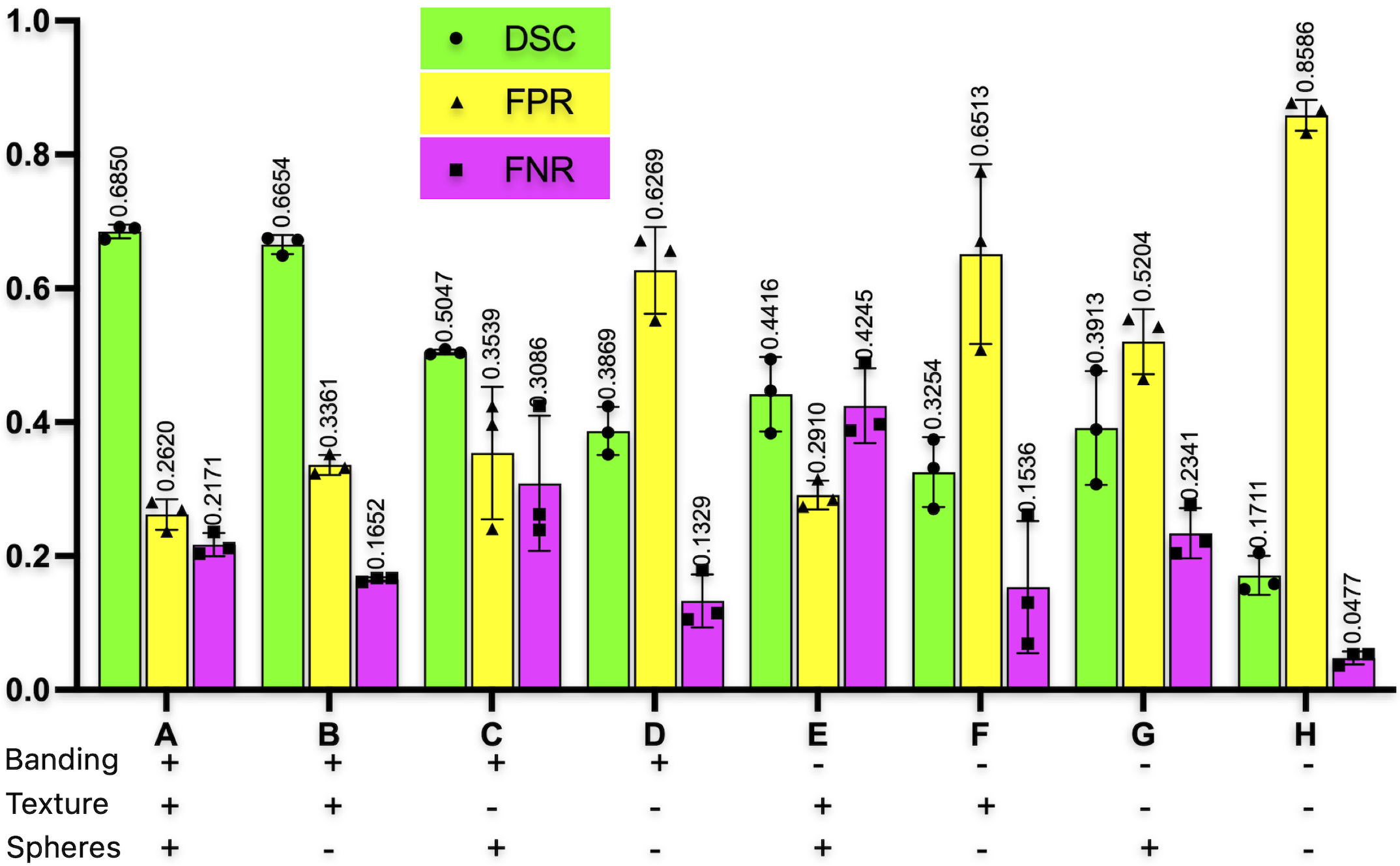}
    \caption{DSC, FPR, and FNR for synthesis conditions A-H (n=3) on human somatosensory cortex with respect to expert human labeler. Mean values shown at top of bar, error bars represent standard deviation.}
    \label{fig:ablation_graph}
\end{figure}

Fig \ref{fig:ablation_images} illustrates the pattern and distribution of true positives (TP) (green), false positive (FP) (yellow), and False negatives (FN) (magenta), for volumetric (subfig (a)) and slice (subfig (b)) predictions of our full model (A) and two increasingly ablated conditions (E and H). Models trained on data synthesized via the proposed method (A) show good agreement with expert labels; however, differences arise in conditions of low CNR, in which our model segments more vessels and maintains better continuity across sections compared to expert labels (blue arrows). Removal of all noise models (H) results in severe FPs, manifesting as banding artifacts (orange arrow). Conversely, the intermediate model (E) is overly conservative, missing many vessels of various sizes almost entirely (red arrows). This trend is illustrated quantitatively in Fig \ref{fig:ablation_graph}, which shows that models trained on datasets with more noise yield higher average DSCs.

\begin{table}[t!]
\centering
\caption{Parameter Estimates from MLR on DSC. $R^2 = 0.8680$}
\label{tab:mlr-dice}
\begin{tabular}{@{}lccccc@{}}
\toprule
Variable & Est. & 95\% CI & $|t|$ & $ p $-value & Sig. \\
\midrule
Intercept & 0.1902 & [0.135, 0.246] & 7.131 & $<0.0001$ & **** \\
z-decay   & 0.2281 & [0.173, 0.284] & 8.553 & $<0.0001$ & **** \\
texture   & 0.1659 & [0.110, 0.222] & 6.218 & $<0.0001$ & **** \\
balls     & 0.1185 & [0.063, 0.174] & 4.441 & 0.0003 & *** \\
\bottomrule
\end{tabular}
\centering
\caption{Parameter Estimates from MLR on FPR. $R^2 = 0.8530$}
\label{tab:mlr-fpr}
\begin{tabular}{@{}lccccc@{}}
\toprule
Variable & Est. & 95\% CI & $|t|$ & $ p $-value & Sig. \\
\midrule
Intercept& 0.8134  & [0.740, 0.887]   & 23.04 & $<0.0001$ & **** \\
z-decay  & -0.1856 & [-0.259, -0.112] & 5.256 & $<0.0001$ & **** \\
texture  & -0.2048 & [-0.279, -0.131] & 5.800 & $<0.0001$ & **** \\
balls    & -0.2614 & [-0.335, -0.188] & 7.403 & $<0.0001$ & **** \\
\bottomrule
\end{tabular}
\centering
\caption{Parameter Estimates from MLR on FNR. $R^2 = 0.5905$}
\label{tab:mlr-fnr}
\begin{tabular}{@{}lccccc@{}}
\toprule
Variable & Est. & 95\% CI & $|t|$ & $ p $-value & Sig. \\
\midrule
Intercept& 0.09971  & [0.0292, 0.170]   & 2.952 & 0.0079        & **   \\
z-decay  & -0.009025& [-0.0795, 0.0614] & 0.2672& 0.7921        & ns   \\
texture  & 0.05930  & [-0.0111, 0.130]  & 1.736 & 0.0945        & ns   \\
balls    & 0.1712   & [0.101, 0.242]    & 5.068 & $<0.0001$   & **** \\
\bottomrule
\end{tabular}
\end{table}

To understand the precise effects of single noise and artefact model ablation, we quantified the impact of each on model prediction quality (DSC, FPR, FNR) even further through multiple linear regression (MLR) (Tables \ref{tab:mlr-dice}, \ref{tab:mlr-fpr}, and \ref{tab:mlr-fnr}, respectively). All regression models were found to be normally distributed, exhibiting a variance inflation factor (VIF) $<$ 1.5.

The DSC regression model was robust, with a $R^2$ value of 0.8680, indicating that 86.80\% of DSC variability is attributable to the predictor variables (banding, intra-vessel texture, and spheres). The model was statistically significant with $p$-value $< 0.0001$ and $F(3,20)=43.84$. The intercept was 0.1902 ($p<0.0001$), suggesting a poor baseline DSC when all noise models are removed. The banding artifact exhibits the greatest positive impact on DSC with a coefficient of $\beta=0.2281$ ($p<0.0001$), with intra-vessel textures and addition of spheres also contributing positively:$\beta=0.1659$ ($p<0.0001$), and $\beta=0.1185$ ($p=0.0003$) respectively.

Table \ref{tab:mlr-fpr} shows the estimation parameters for the FPR model, which also illustrates a strong explanatory power of $R^2=0.8530$. The overall model is statistically significant with $F(3,20)=38.69$, and $p$-value $< 0.0001$. All noise features are considered significant predictors and demonstrate a strong negative effect on FPR.

The regression model for FNR, while statistically significant ($F(3,20)=9.614, p=0.0004$), shows only moderate explanatory power with an $R^2$ of 0.5905, as listed in Table \ref{tab:mlr-fnr}. The only significant predictor of FNR is the sphere artifact, which increases the metric ($\beta=0.1712$, $p<0.0001$).

The best synthesis scheme, therefore is the one including all noise models (A), and will henceforth be used for all future comparisons.

\subsection*{Experiment B. Comparisons with Baseline}
\setcounter{subsubsection}{0}

\begin{figure}
    \centering
    \includegraphics[width=\columnwidth]{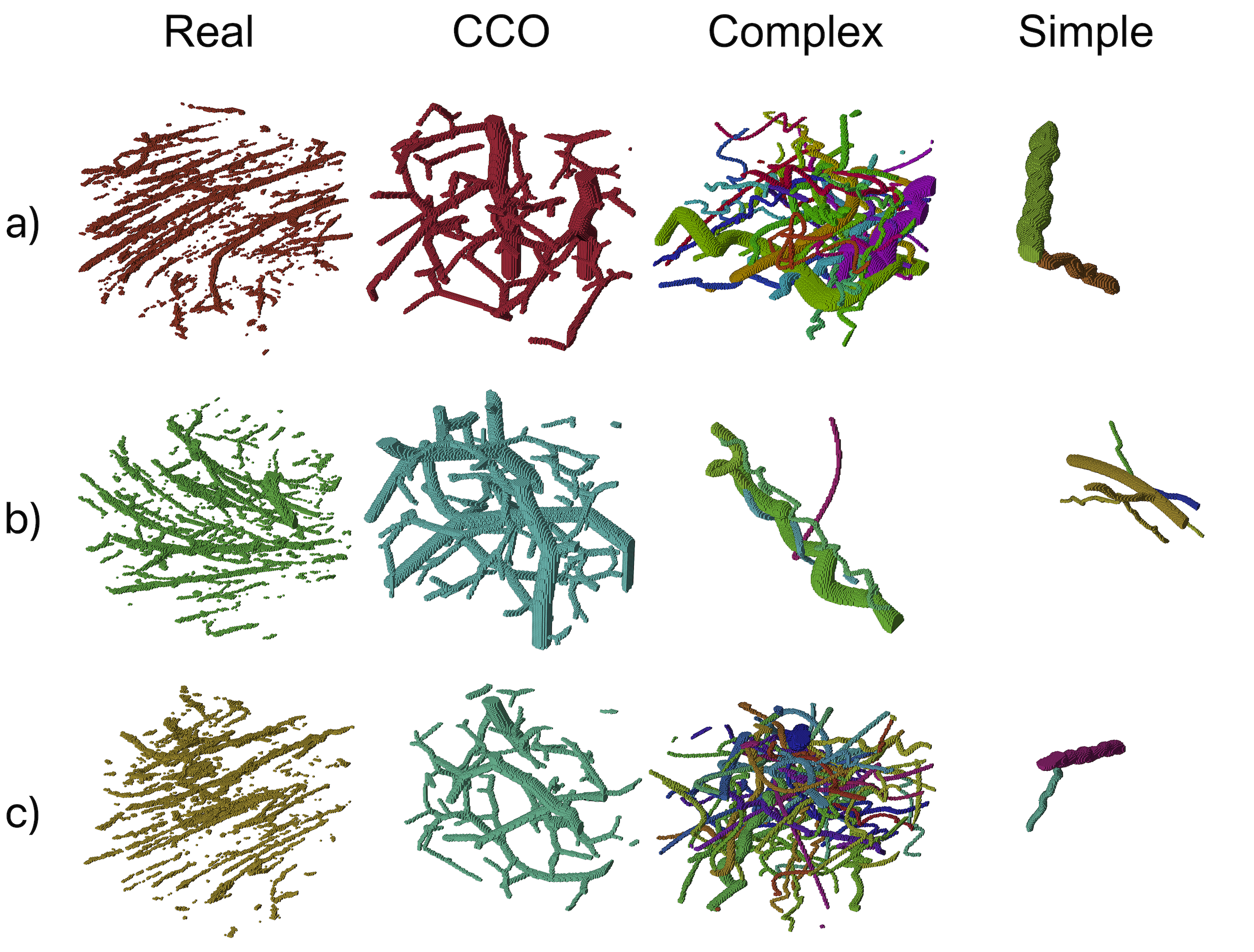}
    \caption{Labels used for training baseline comparisons. Each column represents a different type of label with decreasing realism: 'Real' for actual vascular data, 'CCO' for labels generated via CCO, 'Complex' for synthetic labels with large variance, and 'Simple' for low-variance labels. Rows (a), (b), and (c) display distinct instances, showcasing the variability in complexity and structure of each set.}
    \label{fig:label_types}
\end{figure}

To further assess our synthetic training approach, we evaluated the accuracy of our best model (A) against those trained on more realistic datasets. Each condition serves to examine how training data realism influences segmentation accuracy as measured by DSC. By juxtaposing our best model to several baseline conditions in this way, we aim to better understand the limitations and advantages of our synthetic methods, and whether or not they are detrimental to model performance. The different labels used for training are summarized in Fig \ref{fig:label_types}, and all synthesized intensity volumes follows our comprehensive texture mapping procedure from model A.

\subsubsection{Training Data}

\textbf{sOCT labels \& images (Real+Real)} is an entirely real, no-synth dataset composed of eight $128^3$ voxel patches from the primary somatosensory cortex sample and corresponding label maps by a human expert. The data was split evenly into training and validation sets in a spatially random way. This set resembles the typical standard for modern training sets. The four training patches were augmented by randomly flipping about all three axes to create an augmented set of 32 unique volumes. The quantity of manual labels for sOCT images is scarce, limiting the size of this set.

\textbf{sOCT labels \& synthetic intensity (Real+Synth)} uses the same vascular labels (derived from expert-annotated real data) as Real+Real, but consists of 100,000 unique synthetic intensity volumes from our proposed intensity synthesis engine.

\textbf{CCO Labels \& synthetic intensity (CCO+Synth)}
refers to vessel labels generated via CCO \cite{tetteh2020deepvesselnet}, split into 675 patches of size $128^3$ voxels, textured with the synthesized intensity from our proposed synthetic pipeline to create a training set of 100,000 unique examples. 

\textbf{Complex labels \& synthetic intensity (Complex+Synth)} describes the most complicated situation generated by our proposed high-variance synthesis engine (equivalent to model A from Experiment A). This set consists of 100,000 unique training examples.

\textbf{Simple labels \& synthetic intensity (Simple+Synth)} refers to the 100,000 membered training set generated by a low-variance configuration of our label synthesis engine (simple vessels) and normal intensity synthesis. Specifically, for all parameters sampled in $\log\mathcal{N}(\mu, \sigma^2)$, variances were halved. Similarly, the upper limit ($b$) was halved for parameters sampled in $\mathcal{U}(a, b)$ or $\mathcal{U}_{\mathbb{Z}}(a, b)$.

\subsubsection{Test Data}
Four samples from the frontal and occipital lobes. Two random $64^3$ (1.2 mm$^3$) patches were extracted from each sample's stitched data, yielding eight unique test patches. Two independent labelers annotated each patch exhaustivley for vasculature. These patches formed the held-out test set for final model evaluation and baseline comparison.

\begin{figure*}[!t]
    \centering
    \includegraphics[width=0.9\textwidth]{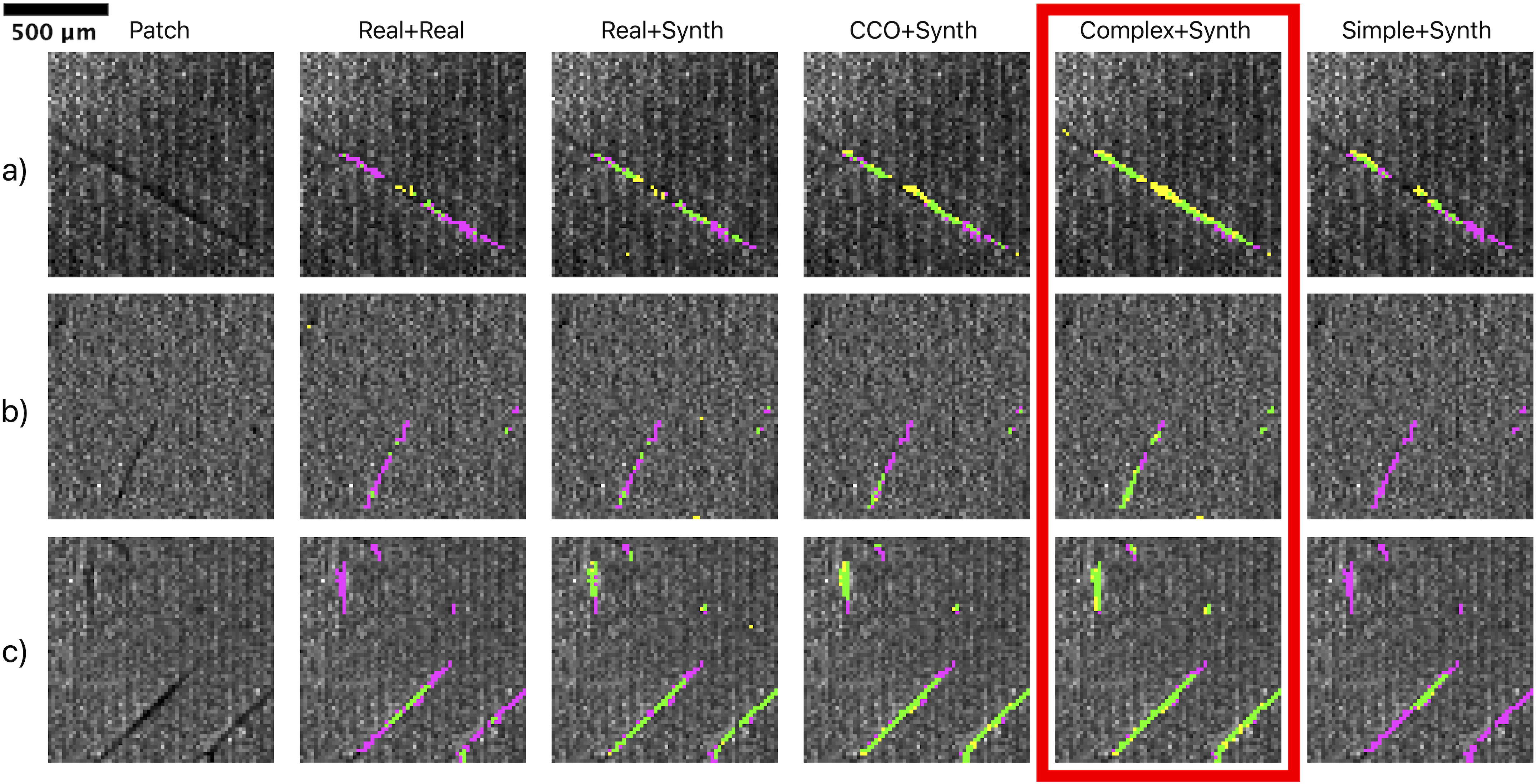}
    \caption{Frontal \& occipital (test) results. Comparative visualization of model predictions across training conditions varying in data realism. Each row shows a unique sample from the test dataset. This figure illustrates the spatial distribution of TPs (green), FPs (yellow), and FNs (magenta) for each condition. Proposed model (Complex+Synth) is indicated by red bounding box. Scale bar (500 {\textmu}m) is shown in upper left corner.}
    \label{fig:baseline_comp_images}
\end{figure*}

\subsubsection{Results}
Fig \ref{fig:baseline_comp_images} illustrates the distribution of TP, FP, and FN, from random slices of different patches of the test set to highlight the impact of data realism on qualitative model performance. The models trained on Real+Real and Simple+Synth almost completely miss the vessels in patch (b), where the vessel contrast is low, and demonstrate a high quantity of FNs even in vessels with high CNR (a and c). In fact, the model trained on Real+Real almost completely misses the large vessel in patch (a). Models trained on Real+Synth show a modest improvement in prediction quality, especially with the increase in TPs on patch (c). Still, this model performs poorly on patch (b). The models trained on CCO+Synth and Complex+Synth show great improvements to prediction quality. These models provide more continuous vessel segmentation across slices. Especially, the Complex+Synth model could show more continuity of vessel segmentation than the human labeler in certain regions across slices where vessel contrast changes over depth (presented by FPs on patch (a)).

\begin{figure}
    \centering
    \includegraphics[width=\columnwidth]{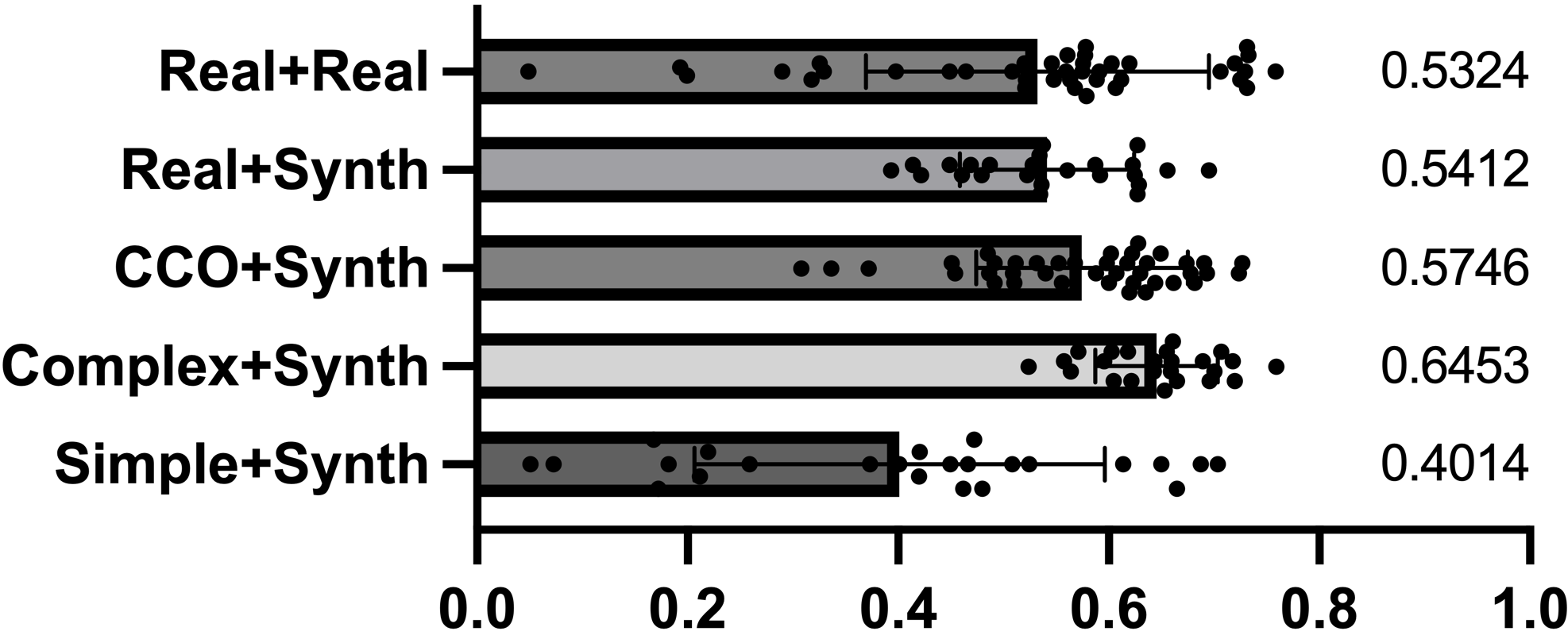}
    \caption{Frontal and occipital (test) results. DSC from baseline training conditions and synthetic condition A (Complex+Synth). DSCs are computed over 8 patches and averaged between to two expert labelers. A one-way ANOVA test with respect to Complex+Synth shows that all conditions are significantly different, except for CCO+Synth.}
    \label{fig:baseline_comp_graph}
\end{figure}

In order to take into account the difference in ground truth labels across raters and the presence of repeated patches across methods, we applied a mixed-effect model to our data, with rater and patch as mixed effects and model as fixed effect \cite{bates2015lme4}. After model fitting, Tukey's range test was applied to the estimated marginal means. Analyses were performed in R using the \texttt{lme4} and \texttt{emmeans} packages. All pairwise comparisons yielded statistically significant differences, except CCO+Synth \emph{vs.} Real+Synth and Real+Synth \emph{vs.} Real+Real. Notably, the proposed model obtained a better DSC than the model trained with CCO labels ($p<0.01$), the model trained with real labels and synthetic images ($p < 0.0001$) and the model trained with real labels and images ($p < 0.0001$). The DSCs for each condition are plotted in Figure \ref{fig:baseline_comp_graph}.

\subsection*{Experiment C. Segmentation Reliability}
\setcounter{subsubsection}{0}

In this experiment, we show that our method achieves segmentation precision statistically comparable to human experts across four distinct imaging conditions.

\subsubsection{Setup}
We randomly selected one of the three runs from our proposed model Complex+Synth to examine our proposed method's reliability with respect to the reliability of human raters. Segmentation results were compared between Human rater 1 (R1), human rater 2 (R2), and a our proposed model (Complex+Synth/A) on each patch from the Test Data of Experiment 2. The Cohen’s kappa coefficient was calculated for each permutation of pairs (R1-R2, R1-A, R2-A) and partitioned into human-human (R1-R2) and human-method (R1-A, R2-A) groups. The Wilcoxon matched-pairs signed rank test was used to compare Cohen's kappas between groups.

\subsubsection{Results}
The Cohen’s kappa coefficient for any pair of raters was moderate, with the highest agreement occurring between the human raters ($\kappa=0.6848 \pm 0.0930$). A Wilcoxon signed-rank test was used to compare the ratings between different pairs of raters. After  correcting for
multiple comparisons (corrected risk $\alpha=0.025$), no statistically significant difference was found between R1-A and R1-R2 (p=0.6406) or R2-A and R1-R2 (p=0.0391), suggesting that automated and human labels cannot be distinguished from each other.

\subsection*{Experiment D. Full Prediction on Human Primary Somatosensory Cortex}
\setcounter{subsubsection}{0}

Our culminating experiment entails full-scale inference on the entirety of the primary human somatosensory cortex sample ($3.4 \times 2.9 \times$ 1.1 cm$^3$, $1716 \times 1470 \times 561$ voxels) shown in Fig. \ref{fig:I46_ref} to evaluate the practicality of our model in a large-scale, real-world application. We show that the model trained on our proposed synthetic data can be applied to a large tissue dataset with minimal preprocessing and no qualitative/noticeable breakdown in accuracy. 

\subsubsection{Setup}
We randomly selected one of the three runs from the Complex+Synth model to predict on the entirety of the primary human somatosensory cortex sample. No additional image processing was used other than creating a tissue mask on the sOCT volume.

\subsubsection{Results}
\begin{figure*}[ht!]
    \centering
    \includegraphics[width=\textwidth]{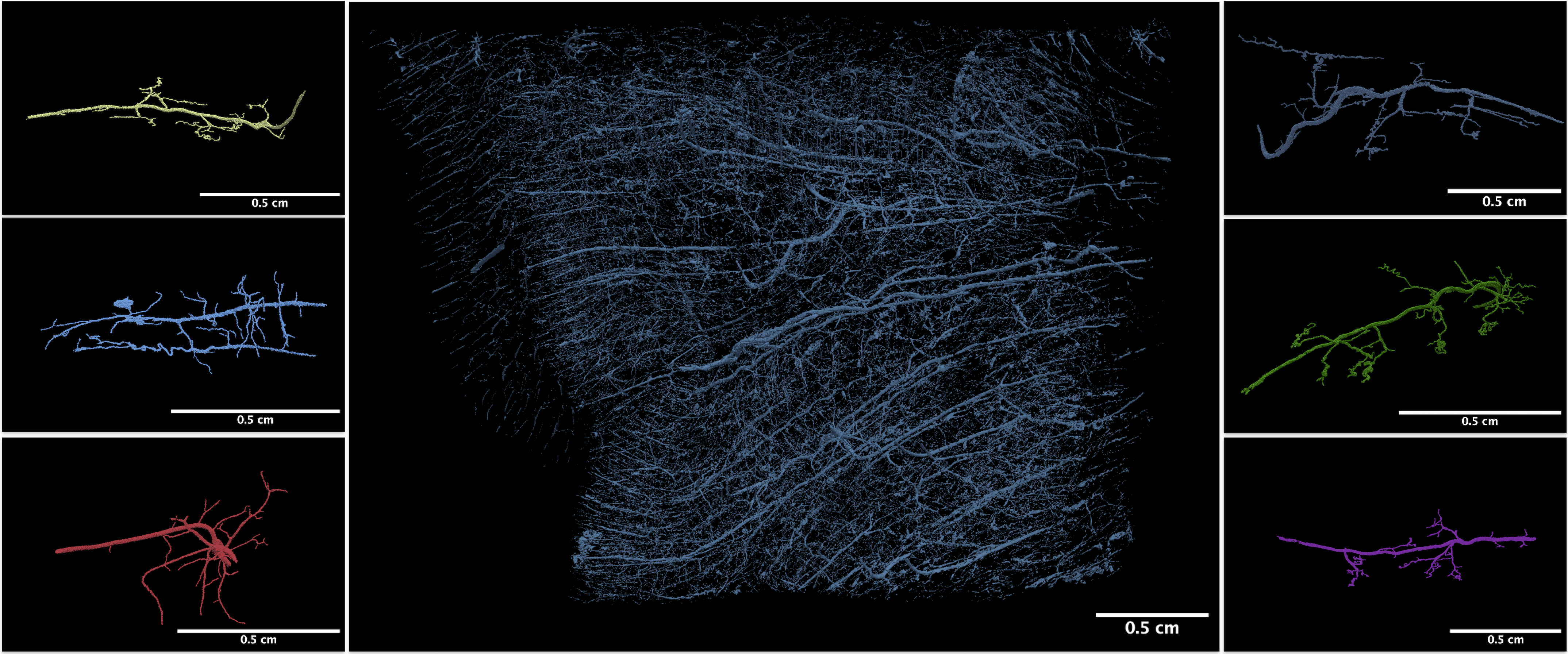}
    \caption{Results of comprehensive vasculature segmentation of human somatosensory cortex volumetric data. Six side panels display the results of connected component extractions, each illustrating a distinct subset of the segmented vascular structures. These extractions highlight specific features and connections within the overall network, providing deeper insights into the spatial distribution and connectivity of vascular components. Scale bars (0.5 cm) shown in white in bottom right corner of each image.}
    \label{fig:connected_components}
\end{figure*}

Fig. \ref{fig:connected_components} shows a volumetric rendering of the full-scale prediction and the six largest eight-connected components by volume. We illustrate a diverse set of continuous, vascular configurations identified by large-scale application of our proposed method. It is evident that our model identifies both highly tortuous vessels and linear ones. It also displays efficacy in identifying vessels that occur at multiple spatial scales, from those whose diameter is a single voxel, to much larger vessels running through the cortex and the white matter.

\section{Discussion}

This study demonstrates that U-Nets are capable of accurately segmenting vessels in large volumetric sOCT images at microscopic resolution by learning vascular features from entirely synthetic datasets.

Despite the enriched information of the vasculature revealed by sOCT, the reconstructed volume presents challenges for vessel segmentation due to high-contrast speckle noise, banding artifacts, and non-vessel structures caused by various tissue scattering. We began by probing the effects of three imaging artifacts (intensity banding, intra-vessel texture, and spheres) on model accuracy (DSC), and found that including each in the augmentation is essential for optimizing the quality of training data for segmenting vasculature in sOCT data. We have illustrated both qualitatively and quantitatively that the specific distribution of FPs and FNs align with what is expected when certain features in the intensity synthesis are ablated. Most notably, the numerous missing vessels when the intensity banding artifact is ablated from the synthetic datasets and the substantial increase in banding-related FPs when the multiple artifact components are removed. These findings underscore the importance of artifact simulation in the creation of synthetic data for model training.

Through comparative analysis of models trained on synthetic and real data, we also note several critical advantages of using the former over the latter. Practically, despite its upfront time investment for tasks such as texture mapping and label synthesis, generating training data proves more efficient compared to annotating real images. For instance, producing one hundred synthetic labels (about 2 hours, not including development time) is considerably faster and less labor-intensive than acquiring (days to weeks) and manually labeling (months to years) a similar number of real images.
We demonstrated that the limited availability of real sOCT training data only leads to poor generalization on unseen test data with different experimental/acquisition settings (Fig \ref{fig:baseline_comp_graph}). It is worthwhile mentioning that the limited training data size, and shifts in image features between the (real) train and test set are two potential contributing factors in this poor performance. In contrast, our proposed synthesis effectively trains a model with good generalizability without bottlenecking the sOCT imaging pipeline. We have also observed a non-negligible inter-rater variability in expert-annotated masks even for small patches ($64 \times 64 \times 64$ voxels), suggesting manual annotation of vessels is error-prone, therefore making it challenging to obtain a reliable ground truth. Synthetic images provide an attractive solution to this, as by construction the labeling of the synthetic images is always completely error-free.

While CCO-based vessel labels already exist and are readily available (for certain imaging resolutions), models trained on them produce less accurate results compared to the labels generated by our method. As a distinct modeling procedure, our method does not rely on the extensive priors necessary for building CCO-based vascular models. With real and CCO-based training sets at our disposal, we show that models trained on our most complex and unrealistic data perform the best with respect to quantitative metrics such as DSC, and qualitative ones of TP, FP, and FN patterns. We observe that our proposed model tends to yield an apparent over-segmentation of vessels compared to human labelers in some regions---yet closer investigation reveals that many supposed FPs actually align with the continuous vessel geometry. As such, many voxels categorized as FPs may, in fact, be real vessels, reflecting the difficulty of generating exhaustive manual annotations. Despite the DSC-depressing effect of this ``oversegmentation,'' these models out-perform their real-data-trained counterpart by an approximate 12\% increase in DSC.

Finally, we show that our models are capable of predicting large scale sOCT volumes without re-tuning parameters, such that the extraction of the vasculature from the entire somatosensory cortex sample of $3.4 \times 2.9 \times 1.1$ cm$^3$ at 20 {\textmu}m resolution ($1716 \times 1470 \times 561$ voxels) takes only 1.5 hours (with 64 averages) on an NVIDIA A6000. We have also demonstrated that our optimized segmentation model achieves prediction variability comparable to two independent human raters across four sOCT acquisitions, representing a departure from the limitations of traditional automated methods such as hessian-based filters, which are highly variable to small domain shifts \cite{yousefi2015segmentation,longo2020assessment,yang2022volumetric}. Our findings echo those of the \texttt{Synth} techniques by showing that models trained on entirely synthetic data are robust and accurate. The high throughput of our models within high resolution sOCT workflows will increase our understanding of the intricate cerebrovascular networks. Future endeavors will aim to augment the current training framework by incorporating new noise models and synthesis regimens, potentially offering insights in bolstering its utility in multiple imaging modalities.

\bibliographystyle{ieeetr}
\bibliography{refs}
\end{document}